\documentclass[a4paper,11pt]{article}
\usepackage{pos}
\usepackage{physics}
\usepackage{bm}
\usepackage[compat=1.1.0]{tikz-feynhand}
\usepackage[subrefformat=parens]{subcaption}
\usepackage{bxtexlogo}

\title{Theoretical study of the $\Sigma N$ cusp in the $K^-d\rightarrow\pi\Lambda N$ reaction}

\author*[a]{Shunsuke Yasunaga}
\author[a]{Daisuke Jido}

\affiliation[a]{Department of physics, Institute of Science Tokyo\\
  2-12-1, Ookayama, Tokyo, Japan}

\abstract{The $K^-d\rightarrow\pi\Lambda N$ reaction is useful for exploring the hyperon-nucleon interaction through final state interactions. In particular, the cusp structure of the $\Lambda N$ invariant mass spectrum at the $\Sigma N$ threshold contains information about the s-wave interaction of 1/2-isospin hyperon-nucleon systems. The calculation of the spectrum is performed with the aim of extracting the scattering length of the $\Sigma N(I=1/2)$ channel that couples to the $\Lambda N$ channel from this reaction, and the results are discussed in comparison with experimental data to highlight the factors that should be considered.}

\FullConference{10th International Conference on Quarks and Nuclear Physics (QNP2024)\\
8-12 July, 2024\\
Barcelona, Spain\\}

\begin{document}
\maketitle
\section{Introduction}

The $K^-d\rightarrow\pi\Lambda N$ reaction serves as a valuable tool for investigating the interaction between hyperons ($Y$) and nucleons ($N$) through extracting information about their final-state interactions. One particular interest is the "$\Sigma N$ cusp" which appears as a sharp feature in the $\Lambda N$ invariant mass spectrum at the $\Sigma N$ threshold. By analyzing this reaction, we expect to obtain information on the $\Sigma N-\Lambda N$ conversion interaction near the $\Sigma N$ threshold, including parameters such as the scattering length. This conversion interaction is essential to the elementary process of the $YN$ interaction, as the one-pion exchange diagram for $\Lambda N\rightarrow\Lambda N$ is forbidden by isospin conservation.

There have been several studies on the $\Sigma N$ cusp phenomenon in the $K^-d\rightarrow\pi^-\Lambda p$ reaction, including one that utilizes experimental data from the $K^-N\rightarrow\pi Y$ reaction~\cite{dalitz1981}.
Observations have been made for the case of in-flight kaons~\cite{braun1977}. Currently, J-PARC E90 experiment~\cite{ichikawa2022} is ongoing, and it is expected to yield data with higher resolution and statistics than previously available. In experimental side, not only data from the $K^-d\rightarrow\pi^-\Lambda p$ reaction, but also the $\pi^+d\rightarrow K^+\Lambda p$~\cite{pigot1985} and $pp\rightarrow K^+\Lambda p$ reactions~\cite{budzanowski2010}, have been measured to examine the $\Lambda p$ invariant mass spectra near the $\Sigma N$ threshold.

In this paper, we present the results of the calculations for the $\Lambda N$ invariant mass spectrum near the $\Sigma N$ threshold in the $K^-d\rightarrow\pi\Lambda N$ reaction. Here, we rake the s-wave scattering length of the $\Sigma N$ system for the spin-triplet state $a_{\Sigma N(I=1/2)}^t$ as a parameter and examine how variations in this parameter affect the shape of spectrum.

\section{Formulation}
The $\Lambda N$ invariant mass spectrum of the $K^- d \rightarrow \pi \Lambda N$ reaction is calculated by integrating the square of its scattering amplitude $|\mathcal{T}|^2$ for kinematic variables as
\begin{equation}\label{equ:CS_formula}
\frac{d\sigma}{dM_{\Lambda N}} =
\frac{M_d M_\Lambda M_N}{(2\pi)^5 4 k_\text{cm} E_\text{cm}^2}
\int|\mathcal{T}|^2 |\bm{p}_\pi| |\bm{p}_\Lambda^*| d\Omega  d\Omega^*,
\end{equation}
where $M_d$, $M_\Lambda$, and $M_N$ are the masses of deuteron, $\Lambda$-baryon, and nucleon, respectively. Kinematical quantities defined in the center of mass frame are the initial momentum $k_\text{cm}$, the total energy $E_\text{cm}$, the momentum of the pion $\bm{p}_\pi$, and the solid angle of the pion $\Omega$. Those in the $\Lambda N$ rest frame are the momentum of the $\Lambda$-baryon $\bm{p}_\Lambda^*$, and the solid angle of $\Lambda$-baryon $\Omega^*$.


\begin{figure}[htbp]
    \begin{minipage}{0.5\linewidth} 
    \begin{center}
    \begin{tikzpicture}[scale=1.0] 
    \begin{feynhand}    
      \vertex [particle] (i1) at (-3,1) {$K^-$};
      \vertex [particle] (i2) at (-3,0) {$N$};
      \vertex [particle] (i3) at (-3,-1) {$N$};
      \vertex [ringblob] (x1) at (-1,0) {$T_{MB}$};
      \filldraw [fill=black] (-2.6,0.2) rectangle (-2.3,-1.2);
      \propag [chasca] (i1) to (x1);
      \propag [fer] (i2) to (x1);
      \vertex [grayblob] (y1) at (1,-0.5) {$T_{YN}$};
      \vertex [particle] (f1) at (3,1) {$\pi^-$};
      \propag [chasca] (x1) to (f1);
      \propag [fer] (x1) to [edge label= $\Sigma$] (y1);
      \propag [plain] (i3) to (-2.3,-1);
      \propag [fer] (-2.3,-1) to (y1);
      \vertex [particle] (f2) at (3,0) {$\Lambda$};
      \vertex [particle] (f3) at (3,-1) {$p$};
      \propag [fer] (y1) to (f2);
      \propag [fer] (y1) to (f3);
    \end{feynhand}
    \end{tikzpicture}
    \end{center}
    \end{minipage}
    \begin{minipage}{0.49\linewidth} 
        \centering
        \caption{Feynman diagram of the foreground ($\Sigma$ exchange) process for the $K^-d\rightarrow\pi^-\Lambda p$ reaction. The thick line connecting nucleons represents the deuteron. In the diagram, $T_{MB}$ denotes the meson-baryon amplitude and $T_{YN}$ denotes the hyperon-nucleon amplitude.}
        \label{fig:fore}
    \end{minipage}
\end{figure}
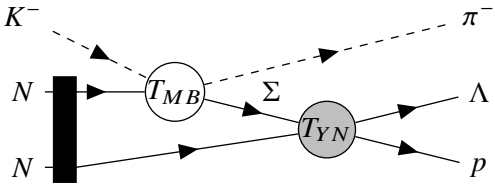

The scattering amplitude $\mathcal{T}$ is formulated as a factorization of up to a two-step amplitude, involving either meson-baryon or hyperon-nucleon amplitudes. Figure~\ref{fig:fore} shows the foreground process, which includes the $\Sigma N\rightarrow\Lambda N$ conversion interaction in the final state. We express the amplitude for this foreground process $\mathcal{T}^{(\Sigma)}$ as
\begin{equation}
\mathcal{T}^{(\Sigma)}(M_{\Lambda N}, \Omega) = \mathcal{N}\ T_{\Sigma N\rightarrow\Lambda N}(M_{\Lambda N})\ T_{K^-N\rightarrow\pi\Sigma}(M_{\Lambda N},\Omega)\ S^\dag \ T_{G}(M_{\Lambda N},\Omega),
\end{equation}
where $\mathcal{N}$ is the normalization factor, $T_{\Sigma N\rightarrow\Lambda N}$ denotes the second-step amplitude for the $\Sigma N\rightarrow\Lambda N$ conversion interaction, $T_{K^-N\rightarrow\pi\Sigma}$ denotes the first-step amplitude for the $K^-N\rightarrow\pi\Sigma$ interaction, and $S^\dag$ is the spin matrix representing the spin-1 deuteron, and $T_G$ denotes the integral of the intermediate $\Sigma$ propagator and the deuteron wave function in momentum space for the s-wave component $\tilde\varphi$, which defined $\psi_0^a$ in CD-Bonn potential~\cite{machleidt2001} as
\begin{equation}
{T}_G =
\int \frac{d^3\vec{q}}{(2\pi)^3} \frac{2M_\Sigma}{q_0^2 - |\vec{q}|^2 - M_\Sigma^2 + i\epsilon} \tilde{\varphi}(|\vec{q} + \vec{p}_\pi - \vec {p}_{K^-}|). \label{equ:tg}
\end{equation}
The energy $q_0$ is assumed to conserve non-relativistic energy with nucleon from the deuteron, given by
\begin{equation}
q_0 = E_{K^-} + M_d - \left(M_N + \frac{|\vec{q}+\vec{p}_\pi-\vec{p}_{K^-}|^2}{2M_N}\right) -E_\pi.
\end{equation}
Not only the foreground process, background processes should also be included such as the $\Lambda$ exchange diagram similar to the foreground, one-step diagram involving only $K^-N\rightarrow\pi\Lambda$ conversion, and meson exchange diagrams; detailed formulations are provided in~\cite{iizawa2022,yasunaga2024}.

For the meson-baryon amplitude $T_{K^-N\rightarrow\pi\Sigma}$, we employ the partial-wave amplitudes up to p-wave ($l=1$) in the strangeness $S=-1$ sector given by Ref.~\cite{jido2002}. We calculate the two-channel amplitude $T_{YN}$ for $\Lambda N$ and $\Sigma N$ using the real symmetric kernel $V$ and momentum matrix $P$ based on the unitarity of the $S$-matrix as
\begin{equation}
T_{YN} = \begin{pmatrix}T_{\Lambda N}&T_{\Lambda N\rightarrow\Sigma N}\\T_{\Sigma N\rightarrow\Lambda N}&T_{\Sigma N}\end{pmatrix} = (-V -iP)^{-1},
\end{equation}
where momentum matrix $P$ is constructed by each center-of-mass momentum $P_{
\Lambda N}$ and $P_{\Sigma N}$ as
\begin{equation}
P = \begin{pmatrix}P_{\Lambda N}&0\\0&P_{\Sigma N}\end{pmatrix}.
\end{equation}
The constant matrix $V$ is fixed so that it takes the empirical value $a_{\Lambda p}^t = -1.56^{+0.19}_{-0.22}$ fm~\cite{budzanowski2010} for $T_{\Lambda N}$ component at the $\Lambda N$ threshold and the theoretically calculated values $a^t_{\Sigma N(I=1/2)}$ by NSC97f~\cite{rijken1999}, J\"ulich'04~\cite{haidenbauer2005}, and SMS~N$^2$LO~(600)~\cite{haidenbauer2023} for $T_{\Sigma N}$ component at the $\Sigma N$ threshold. The model configuration is described in the appendix of Ref.~\cite{iizawa2022}. The pair of values for $a^t_{\Sigma N(I=1/2)}$ and the scattering length for $\Sigma N\rightarrow\Lambda N$ conversion in this model $a^t_{\Sigma N\rightarrow\Lambda N}$, which means the value of off-diagonal component at the $\Sigma N$ threshold, are given in Table~\ref{tab:param}.

\begin{table}[htbp]
    \begin{minipage}{0.54\textwidth} 
        \centering
        \begin{tabular}{c|cc}
        &$a^t_{\Sigma N(I=1/2)}$ [fm]&$a^t_{\Sigma N\rightarrow\Lambda N}$ [fm]\\\hline
        Param.~1&$1.68-2.35i$~\cite{rijken1999}  &$1.28-0.03i$\\
        Param.~2&$-3.83-3.01i$~\cite{haidenbauer2005} &$0.24-1.43i$\\
        Param.~3&$2.53-2.64i$~\cite{haidenbauer2023} &$1.33+0.25i$
        \end{tabular}
    \end{minipage}
    \begin{minipage}{0.45\textwidth} 
        \caption{Parameters for the hyperon-nucleon amplitude at the $\Sigma N$ threshold. The spin-triplet scattering lengths for $\Sigma N(I=1/2)$ system $a^t_{\Sigma N(I=1/2)}$ are the input parameters for our calculation and that for the conversion $a^t_{\Sigma N\rightarrow\Lambda N}$ are corresponding calculated result at the threshold.}
        \label{tab:param}
    \end{minipage}
\end{table}

\section{Results and duscussion}
We describe the results for the shape of the $\Lambda p$ invariant mass spectrum near the $\Sigma N$ threshold obtained from calculations for each parameter listed in Table~\ref{tab:param}. The goal here is to observe the variation in the shape of the spectrum when the parameters are varying. 

Here, we present the results that include the effects of background processes. To suppress the backgrounds, it is sufficient to select a phase space where the pion is emitted forward direction and the nucleon has relatively high momentum. This has been recognized, and also be confirmed in our previous work~\cite{yasunaga2024}. The fact that the $\Sigma N\rightarrow\Lambda N$ interaction depends only on the spin-triplet component is due to the pion being emitted forward.

\begin{figure}
    \centering
    \includegraphics[width=0.65\linewidth]{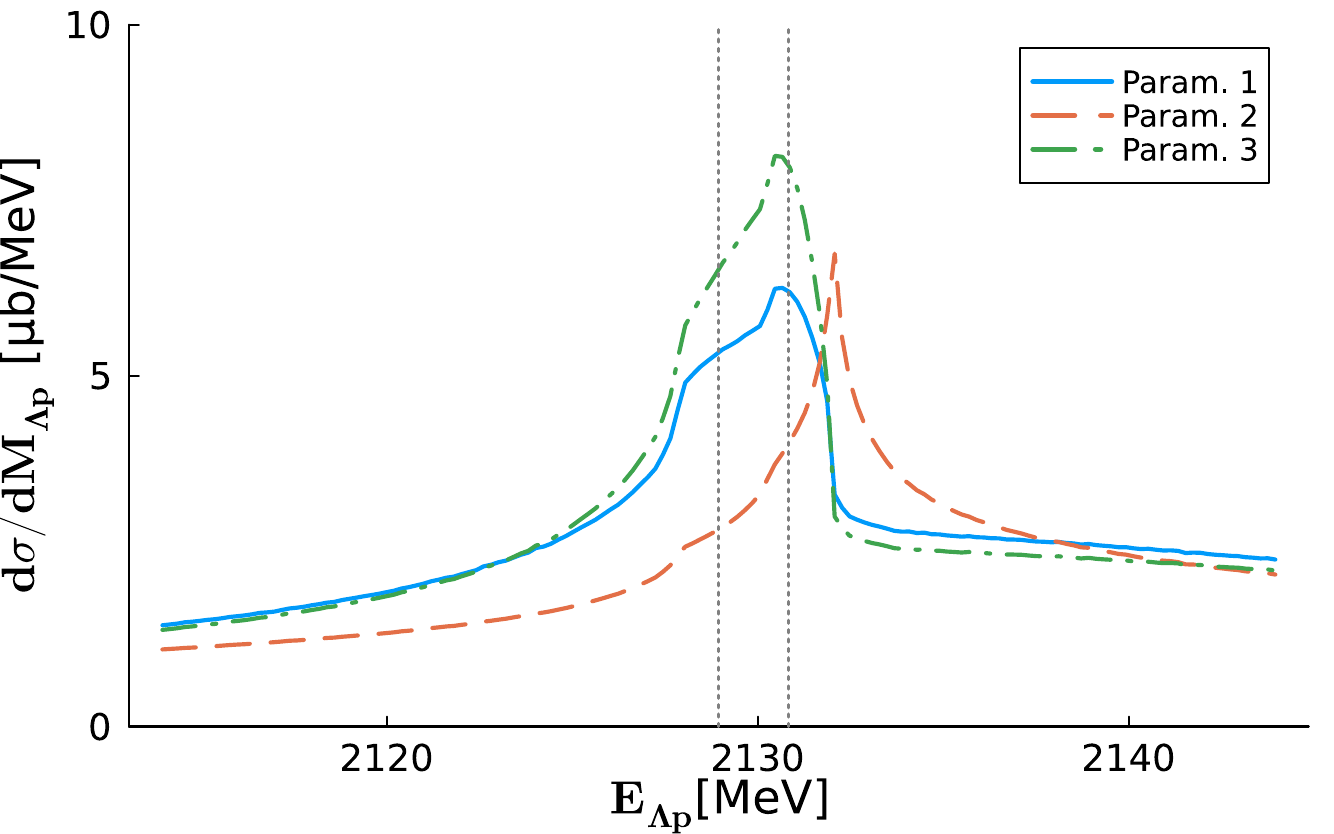}
    \caption{The $\Lambda p$ invariant mass spectrum near the $\Sigma N$ threshold for each parameter. Two vertical dotted lines indicate the two $\Sigma N$ thresholds: $\Sigma^+n$ (2128.9 MeV) and $\Sigma^0p$ (2130.8 MeV), respectively.}
    \label{fig:sigma_cusp}
\end{figure}

Figure~\ref{fig:sigma_cusp} shows the spectrum for each parameter. These results include the contribution from background processes, and we apply kinamatic conditions of $\cos\theta_\pi>0.9$ and $p_p>150$ MeV/c to suppress them. This figure shows the differences in spectrum shapes corresponding to each parameter. The results for Param.~1 and Param.~3 are similar, as expected from the similarity in the values shown in Table~\ref{tab:param}. In particular, cusps can be observed at the $\Sigma^0 p$ threshold. The result for Param.~2 differs from these, with the cusp located to the right of the $\Sigma N$ threshold. Although the results are not shown here, this reflects the structure of the amplitude due to the scattering length near the threshold.

\begin{figure}[htbp]
    \centering
    \includegraphics[width=0.65\linewidth]{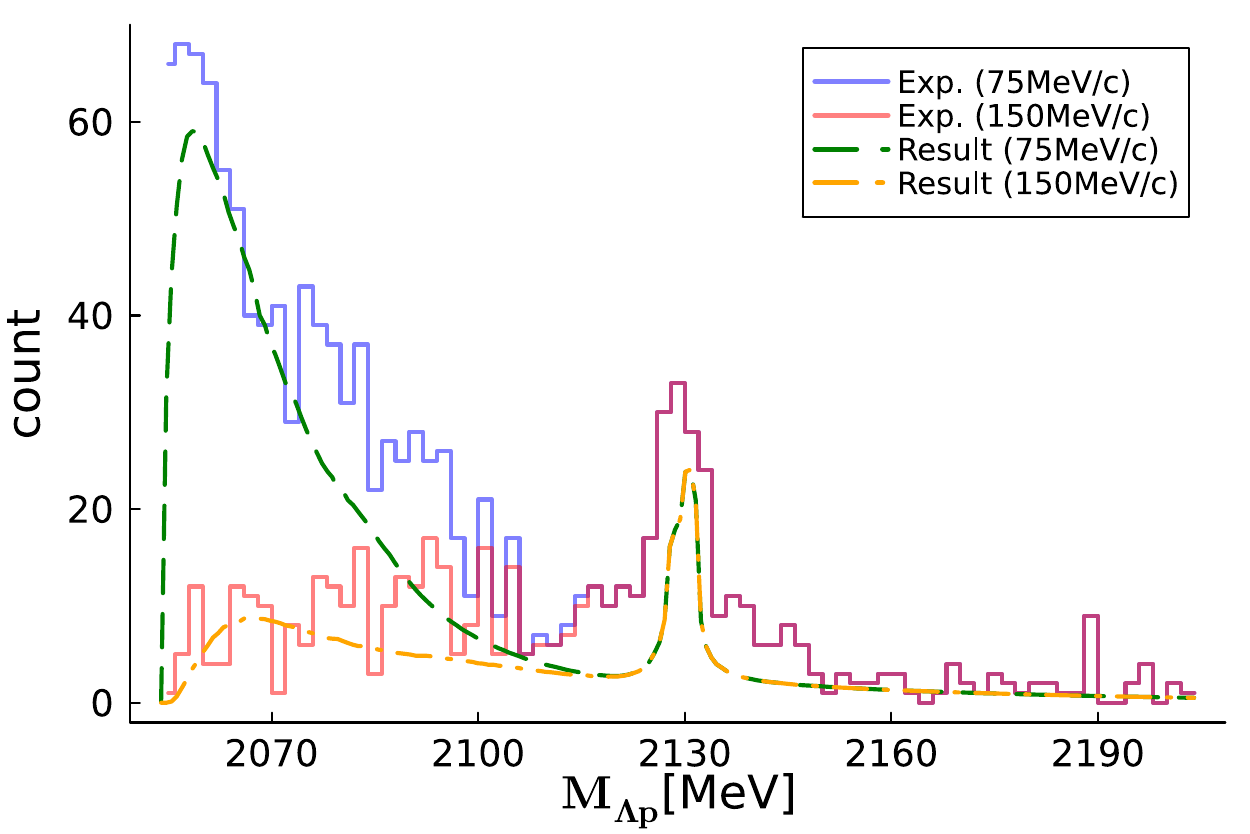}
    \caption{Comparison of our results and experimental data~\cite{braun1977}. The two histograms represent experimental values, with higher one showing the number of events where the final state pion momentum is below 75 MeV/c, and the lower one representing the number of events where it is below 150 MeV/c. Dashed and dash-dotted lines in the figure show our results with corresponding momentum selection of pion, respectively.}
    \label{fig:comp}
\end{figure}

To verify whether the spectra we calculated are realistic, we compare them with an 
old experiment. Here, we focus particularly on whether the backgrounds effects appropriately included. Figure~\ref{fig:comp} shows a comparison between our calculated spectra with Param.~1 and experimental data~\cite{braun1977}. We aim to evaluate the backgrounds, so we take Param.~1 as an example. In the experiment, the momenta of kaons are between 680 and 840 MeV/c, and events are selected where the angle of the pion $\cos\theta_\pi$ is larger than 0.9, with the proton momenta being at least 75 MeV/c or 150 MeV/c. Our calculation results are obtained with the incident kaon having 760 MeV/c momentum and applying the same conditions for the final state as in the experiment. Additionally, our results are scaled by considering the experimental incident flux of 7.1~counts/$\mathrm{\mu}$b.

Near the $\Lambda N$ threshold in Fig.~\ref{fig:comp}, the difference between the low and high momentum cuts is significantly wide, which indicates that the influence of the one-step diagram has been suppressed with the high momentum cut. This is consistently seen both in the theoretical calculation and the experiment. Between the $\Lambda N$ threshold and $\Sigma N$ threshold, the contribution from meson exchange processes become stronger. Even in this region, there are no significant discrepancies. And then, we observe the region near the $\Sigma N$ threshold, which is our main objective. It can be seen that the spread of the tail near the $\Sigma N$ threshold is smaller in our results. In this region, various factors can be considered collectively, $\Sigma N$ scattering length as confirmed in Fig.~\ref{fig:sigma_cusp}, as well as the effective range of the s-wave $\Sigma N$ interaction, the higher partial wave $\Sigma N$ interactions, and other contributing factors that we did not take into account. At the $\Sigma N$ threshold, direct comparison with the experimental data is not possible due to the resolution limitations of the data. Additionally, it is necessary to consider that there are two states at the $\Sigma N$ threshold; $\Sigma^+ n$ and $\Sigma^0 p$. The threshold is split into these two states, and further, due to the isospin symmetry, the ratio of the spectra is expected to be 2:1. In other words, the spectrum is expected to show a tendency to become larger on the left side. In our calculations, we distinguish between the two $\Sigma N$ thresholds, but such a asymmetry is not clearly observed. Due to several factors like this, high-resolution experimental data and accurate calculation are required.

\section{Summary and prospect}
In this study, we calculate the $\Lambda N$ invariant mass spectrum of the $K^-d\rightarrow\pi\Lambda N$ reaction. We focus on the $\Sigma N$ cusp shape near the $\Sigma N$ threshold and discuss the possibility to extract the scattering parameters for $\Sigma N\rightarrow\Lambda N$ conversion interaction. Our results properly evaluate the background process, and it has been confirmed that the shape of the spectrum corresponds to the spin-triplet $\Sigma N(I=1/2)$ scattering length. This suggests that, through comparison with high-resolution experimental data, there is a possibility of indirectly measuring the scattering length.

We plan to update calculations as follows in future. One is to use a $\bar{K}N$ amplitude, which has been accurately fitted to two-body scattering, in order to perform appropriate calculations for higher incident momentum of kaons. Additionally, by incorporating information from higher partial waves, we will perform an accurate comparison of the angular distribution of the spectrum.

\acknowledgments
Attendance of S.Y. at this conference was financially supported by Grants-in-Aid for Scientific Research from JSPS (21K03530).


\begin{thebibliography}{99}


\bibitem{dalitz1981}R. H. Dalitz,
\emph{$\Lambda$- and $\Sigma$-hypernuclear physics},
\href{https://www.sciencedirect.com/science/article/pii/0375947481905959}
{Nuclear Physics A 354, 101 (1981).}

\bibitem{braun1977}O. Braun {\it et al}.,
\emph{On the $\Lambda p$ enhancement near $\Sigma N$ threshold},
\href{https://www.sciencedirect.com/science/article/pii/0550321377902759}
{Nuclear Physics B 124, 45 (1977).}

\bibitem{pigot1985}C. Pigot {\it et al}. (Rome-Saclay-Vanderbilt Collaboration),
\emph{Study of the production of a strange $(S=-1)$ dibaryon in the interactions of $K^-$ and $\pi^+$ on deuterium below 1.5G eV/c},
\href{https://www.sciencedirect.com/science/article/pii/0550321385900458}
{Nuclear Physics B 249, 172 (1985).}

\bibitem{budzanowski2010}A. Budzanowski {\it et al}. (HIRES Collaboration),
\emph{High resolution study of the $\Lambda p$ final state interaction in the reaction $p+p\rightarrow K^++(\Lambda p)$},
\href{https://ui.adsabs.harvard.edu/abs/2010PhLB..687...31H/abstract}
{Physics Letters B 687, 31 (2010).}

\bibitem{ichikawa2022}Y. Ichikawa {\it et al}. (J-PARC E90 Collaboration),
\emph{High resolution spectroscopy of the “$\Sigma N$ cusp” by using the $d(K^-,\pi^-)$ reaction},
\href{https://www.epj-conferences.org/articles/epjconf/abs/2022/15/epjconf_hyp2022_02012/epjconf_hyp2022_02012.html}
{EPJ Web Conf. 271, 02012 (2022).}
\bibitem{machleidt2001}R. Machleidt,
\emph{High-precision, charge-dependent Bonn nucleon-nucleon potential},
\href{https://journals.aps.org/prc/abstract/10.1103/PhysRevC.63.024001}
{Phys. Rev. C 63, 024001 (2001).}

\bibitem{iizawa2022}Y. Iizawa, D. Jido, and T. Ishikawa,
\emph{$K^-d\rightarrow\pi\Lambda N$ reaction for studying charge symmetry breaking in the $\Lambda N$ interaction},
\href{https://journals.aps.org/prc/abstract/10.1103/PhysRevC.106.045201}
{Phys. Rev. C 106, 045201 (2022).}
\bibitem{yasunaga2024}S. Yasunaga, D. Jido, T. Ishikawa, (2024),
\emph{$K^-d\rightarrow\pi\Lambda N$ reaction with in-flight kaons for studying the $\Lambda N$ interaction},
\href{https://arxiv.org/abs/2405.15534}
{[{\tt nucl-th/arxiv:2405.15534}].}

\bibitem{jido2002}D. Jido, E. Oset, and A. Ramos,
\emph{Chiral dynamics of the p wave in $K-p$ and coupled states},
\href{https://journals.aps.org/prc/abstract/10.1103/PhysRevC.66.055203}
{Phys. Rev. C 66, 055203 (2002).}
\bibitem{rijken1999}Th. A. Rijken, V. G. J. Stoks, and Y. Yamamoto,
\emph{Soft-core hyperon-nucleon potentials},
\href{https://journals.aps.org/prc/abstract/10.1103/PhysRevC.59.21}
{Phys.Rev. C 59, 21 (1999).}
\bibitem{haidenbauer2005}J. Haidenbauer and U.-G. Meißner,
\emph{Jülich hyperon-nucleon model revisited},
\href{https://journals.aps.org/prc/abstract/10.1103/PhysRevC.72.044005}
{Phys. Rev. C 72, 044005 (2005).}
\bibitem{haidenbauer2023}J. Haidenbauer, U.-G. Meißner, A. Nogga, and H. Le,
\emph{Hyperon–nucleon interaction in chiral effective field theory at next-to-next-to-leading order}
\href{https://link.springer.com/article/10.1140/epja/s10050-023-00960-6}
{Eur. Phys. J. A 59, 63 (2023).}
\end{thebibliography}
\end{document}